\documentclass[11pt]{article}
\usepackage{moriond}
\pdfoutput=1

\def\Journal#1#2#3#4{{#1} {\bf #2}, #3 (#4)}

\def\PLB{{\em Phys. Lett.}  B}
\def\PRL{\em Phys. Rev. Lett.}
\def\PRD{{\em Phys. Rev.} D}
\def\PR{\em Phys. Rev.}

\def\be{\begin{equation}}
\def\ee{\end{equation}}
\def\bea{\begin{eqnarray}}
\def\eea{\end{eqnarray}}

\newcommand{\met}{\mbox{$\not\!\!E_T$}}
\newcommand{\jpzm}{$J^{P}=0^{-}$}
\newcommand{\jpzp}{$J^{P}=0^{+}$}
\newcommand{\jptp}{$J^{P}=2^{+}$}
\newcommand{\Photo}{}

\begin{document}
\vspace*{4cm}
\title{SPIN AND PARITY IN THE \boldmath{$WH \rightarrow \ell \nu b\bar{b}$} CHANNEL AT THE D0 EXPERIMENT}

\author{ E. JOHNSON }

\address{Department of Physics and Astronomy, Michigan State University\\ 567 Wilson Road, Rm 3220\\
East Lansing, MI 48824 USA}

\maketitle\abstracts{
We probe the spin ($J$) and parity ($P$) of the Standard Model Scalar (SMS) 
using existing searches sensitive to its production rate and kinematic 
properties. In particular, we search for the SMS decaying to a pair of $b$-quarks produced in 
association with a $W$ boson decaying leptonically. By exploiting the differences in kinematic observables, namely the 
transverse mass of the final state products ($M_{T}$), we attempt to distinguish between three 
possible $J^{P}$ hypotheses: the standard model prediction (\jpzp), a 
pseudoscalar (\jpzm), and a graviton-like particle (\jptp). 
With $9.7$~fb$^{-1}$ of data collected at the D0 experiment we show the expected sensitivity to different spin 
and parity hypotheses.
}

\section{Introduction}

Following the discovery of a boson consistent with the predicted Standard 
Model Scalar (SMS)~\cite{higgs,englert} by the ATLAS~\cite{atlas} and CMS~\cite{cms} experiments at CERN 
and the evidence for its decay into two bottom quarks at the Tevatron experiments~\cite{tev}
 it is very important to test its properties. The standard model (SM) predicts that 
the SMS will be spin zero with even parity (\jpzp). The observed 
decay to a pair of photons at the ATLAS experiment eliminates spin one as a possibility according 
to the Landau-Yang Theorem~\cite{landau,yang}. Therefore, the simplest and most 
well-motivated possibilities include the standard model prediction (\jpzp), a spin zero pseudoscalar 
(\jpzm), and a spin two particle with graviton-like couplings (\jptp). 
Although ATLAS and CMS have excluded the \jpzm\ and $2^{+}$ in 
the diphoton and four lepton final states~\cite{atlas_spin,cms_spin} they have yet to study
 the $b\bar{b}$ final state.

Searches for the SMS produced in association with a $W$ or $Z$ boson (henceforth $V$ boson) 
are sensitive to the different kinematics of the three $J^P$ hypotheses. 
This is seen most starkly in the invariant mass of the final state products~\cite{ellis}, $Vb\bar{b}$.
 The mature $VH \rightarrow Vb\bar{b}$ analyses from the D0 experiment are therefore good candidates for study. 
This paper focuses on the $WH \rightarrow \ell \nu b\bar{b}$~\cite{wh} channel specifically though the 
corresponding studies in the $VH \rightarrow \nu \nu b\bar{b}$~\cite{vh} and $ZH \rightarrow \ell \ell b\bar{b}$~\cite{zh}
 channels are underway. We employ the most recent published analysis with no modifications to the 
event selection or analysis methodologies. 

\section{Data and Simulated Samples}

This analysis uses $9.7$~fb$^{-1}$ of data collected by the D0 detector at Fermilab. Our SM 
background samples and SM signal are produced and simulated using {\sc alpgen}~\cite{alpgen}, 
{\sc pythia}~\cite{pythia}, and {\sc singletop}~\cite{singletop} or estimated from data. The \jpzm\ and \jptp\ samples 
were created using {\sc madgraph}$5$ version $1.4.8.4$~\cite{madgraph}. Within the {\sc madgraph} software 
there are several non-SM models as well as the ability to create user-defined models. We follow 
the prescription in Ellis {\it et al.}~\cite{ellis}, using the Randall-Sundrum graviton model 
for the \jptp\ sample and a user-defined model for the \jpzm\ pseudoscalar sample.
 The mass of the SMS was set to $125$~GeV, close to the 
measured value of the discovered boson. The PDF set used in the generation was {\sc cteq6l1}. 
These samples were then showered using {\sc pythia}, reconstructed, and processed through the 
full detector simulation. 

\begin{figure}[htp]
\begin{center}
\includegraphics[width=0.5\linewidth]{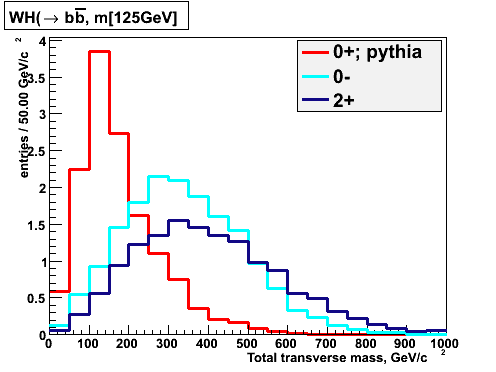}
\end{center}
\caption{(color online) Transverse mass of the $\ell \nu b\bar{b}$ system for the \jpzp\, \jpzm\, and \jptp\ signals. 
These are shown after reconstruction for the two tight tag channel. The histograms are normalized to 
unit area.}
\label{fig:gen}
\end{figure}

\section{Event Selection}

We summarize the event selection for the $WH \rightarrow \ell \nu b\bar{b}$ channel. More details can 
be found in A. Abazov {\it et al.}~\cite{wh}.  We require 
one lepton ($\mu$ or $e$), large missing transverse energy (\met), and two or three jets in the final state. 
We utilize b-tagging to identify the decay products of the SMS and divide the events into 
four orthogonal tagging categories: ``one tight tag,'' ``two loose tags,'' ``two medium tags,'' and 
``two tight tags.'' A Boosted Decision Tree (BDT) is trained for each tagging category 
and jet multiplicity to separate SM signal from background. 

\section{Final Variable Distributions}

In order to differentiate between the three spin-parity assignments we use the transverse mass of the 
$\ell \nu b\bar{b}$ system defined by

\begin{equation}
M_{T}^{2} = \left( E_{T}^{W} + E_{T}^{H} \right)^{2} - \left(\vec{p}_{T}^{W} + \vec{p}_{T}^{H} \right)^{2}
\label{eq:mt}
\end{equation}

\noindent
where the transverse momentum of the $W$ boson, $\vec{p}_{T}^{W}$, is defined as 

\begin{equation}
\vec{p}_{T}^{W} = \vec{\met} + \vec{p}_{T}^{\ell}
\label{eq:pt}
\end{equation}

\begin{figure}[htp]
\begin{center}
\includegraphics[width=0.48\linewidth]{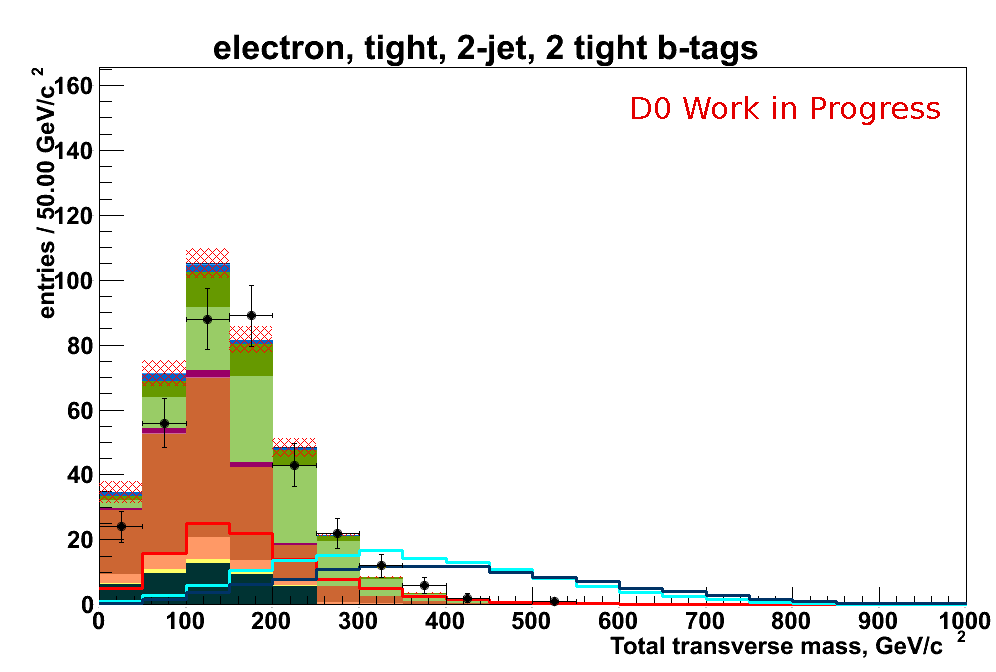}
\includegraphics[width=0.48\linewidth]{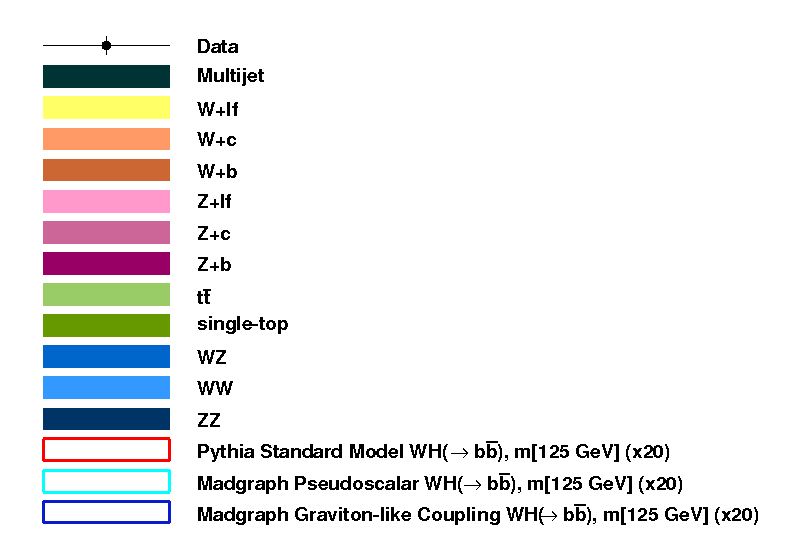}
\end{center}
\caption{(color online) Transverse mass of the $\ell \nu b\bar{b}$ system with accompanying legend.}
\label{fig:mcbkgd}
\end{figure}

\noindent
To illustrate the differences between the three hypotheses for this kinematic variable, the three 
signal hypotheses are shown in Fig.~\ref{fig:gen} after reconstruction. In addition to being 
able to differentiate between the test signals (i.e. the \jpzm\ and $2^{+}$ signals), Fig.~\ref{fig:mcbkgd} shows the non-SM test signals 
peaking in a different region than either the SM signal or the other backgrounds. To ensure adequate 
background statistics when performing our statistical analysis, 
we rebin our $M_{T}$ distribution so that bins above $600$~GeV are 
included as a single overflow bin. 

\section{Statistical Interpretation}

We use a modified frequentist (CL$_{S}$) approach with a log-likelihood ratio test statistic (LLR) for two 
hypotheses: the test hypothesis $H_{1}$ and the null hypothesis $H_{0}$. The LLR test statistic is given by

\begin{equation}
LLR = -2\ln \frac{L_{H_{1}}}{L_{H_{0}}}
\end{equation}

\noindent
where $L_{H_{x}}$ is the likelihood function for the hypothesis $H_{x}$. For a typical SMS search $H_{0}$ is the 
background-only hypothesis and $H_{1}$ is the signal plus background hypothesis. We can view our analysis through 
two different paradigms:

\begin{description}
\item[$\mathbf{H_{0} = b}$ and $\mathbf{H_{1} = s_{1}+b}$] here we decide if the data looks more like it came from a background-only 
distribution or from a test signal plus background distribution
\item[$\mathbf{H_{0} = s_{0}+b}$ and $\mathbf{H_{1} = s_{1}+b}$] here we decide if our data looks more like it came from a SM signal 
 plus background distribution or from a test signal plus background distribution
\end{description}

\begin{figure}[htp]
\begin{center}
\includegraphics[width=0.48\linewidth]{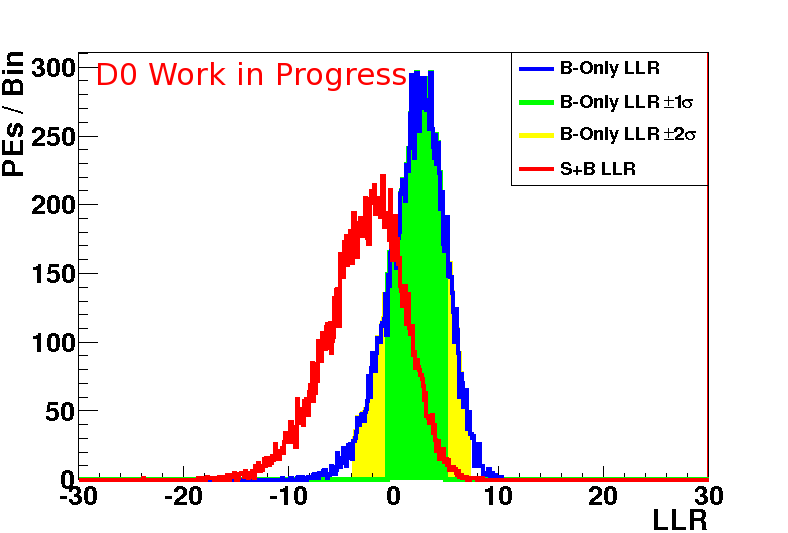}
\includegraphics[width=0.48\linewidth]{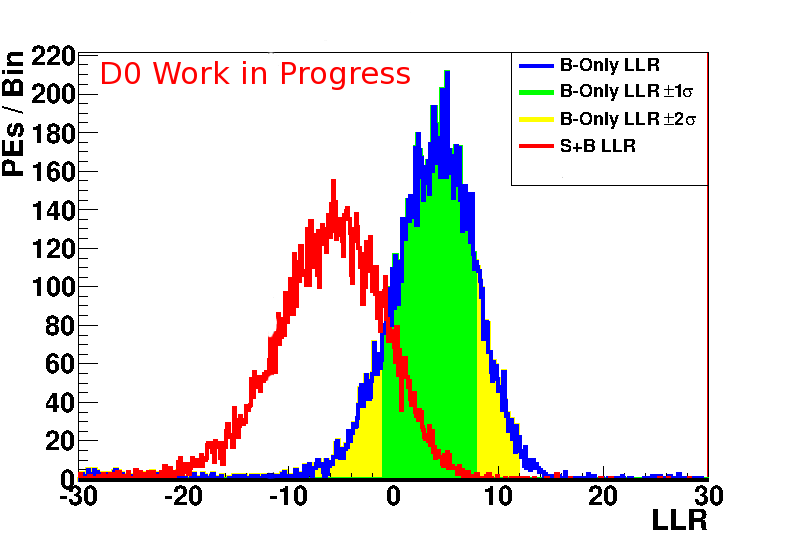}
\end{center}
\caption{(color online) LLR distributions for the $WH \rightarrow \ell \nu b\bar{b}$ channel (left) and the combination of the $VH \rightarrow Vb\bar{b}$ 
channels (right) for the \jptp\ plus background and background-only hypotheses.}
\label{fig:llr}
\end{figure}

\noindent
 To gain insight on the separation 
significance we look at the LLR distributions populated by simulated experiments, assuming Poisson statistics,
 drawn from populations of the test and null hypotheses. 
As a preliminary result, we present the LLR distributions for the \jptp\ plus background hypothesis and background-only hypothesis
for the $WH \rightarrow \ell \nu b\bar{b}$ channel (Fig.~\ref{fig:llr} (left)) and all $VH \rightarrow Vb\bar{b}$ channels 
combined (Fig.~\ref{fig:llr} (right)). The spatial separation between the signal plus background and background-only 
LLR distributions is a good illustration of how effective the analysis is at separating the signal plus background 
and the background-only hypotheses. From these LLR distributions it is clear then that we achieve 
good separation between the signal plus background vs. background-only hypotheses.

\section{Summary}

In summary, we presented preliminary results from the D0 experiment on the spin and parity of the newly observed SMS boson 
in the $WH \rightarrow \ell \nu b\bar{b}$ channel. While presently unable to show the observed LLR line, it is clear that 
we achieve good separation between the hypotheses. 

\section*{Acknowledgments}

We thank the staffs at Fermilab and collaborating institutions,
and acknowledge support from the
DOE and NSF (USA);
CEA and CNRS/IN2P3 (France);
MON, NRC KI and RFBR (Russia);
CNPq, FAPERJ, FAPESP and FUNDUNESP (Brazil);
DAE and DST (India);
Colciencias (Colombia);
CONACyT (Mexico);
NRF (Korea);
FOM (The Netherlands);
STFC and the Royal Society (United Kingdom);
MSMT and GACR (Czech Republic);
BMBF and DFG (Germany);
SFI (Ireland);
The Swedish Research Council (Sweden);
and
CAS and CNSF (China).



\end{document}